\documentstyle[aps,twocolumn]{revtex}

\newfont{\ssrm}{cmr8}
\newfont{\srm}{cmr12}

\newcommand{\bvec}[1]{\mathop{\mbox{\boldmath $#1$}}\nolimits}

\def\Tr{{\rm Tr}}

\begin{document}

\title{Irreversibility from a Reversible Equation}

\author{Hiroshi Ezawa}

\address{Department of Physics, Gakushuin University\\
Mejiro, Toshima-ku, Tokyo 171-8588, Japan}

\author{Koichi Nakamura}
\address{Division of Natural Science, Izumi Campus\\
Meiji University, Eifuku, Suginami-ku, Tokyo 168-8555, Japan}

\author{Keiji Watanabe}
\address{Department of Physics, Meisei University\\
Hino, Tokyo 191-8506, Japan}


\maketitle

\begin{abstract}
After a discussion on the state of local equilibrium
with temperature inhomogeneity, comparing mixture state reprsentation
in statistical mechanics and pure state representation in
thermo field dynamics, a simple model is solved to show that a
reversible equation of motion with the initial condition having
inhomogeneous temperature can lead to irreversible, {\it viz.}
diffusive, behaviour. Yet, the
solution is time symmetric exhibiting diffusion both towards future and
past.
\end{abstract}


\section{INTRODUCTION}

The irreversible processes remain
to be understood despite the long history and many attempts\cite{ich}
\cite{nak}.
In particular, those processes caused by temperature gradient, say, heat
conduction and thermal diffusion, have no approaches from the first 
principles that are commonly accepted,
while those driven mechanically, say, electric conduction, has been
handled by the perturbation theory in Hamiltonian dynamics, at least for
the purpose of computing the transport coefficients\cite{kub}.

We have been trying to obtain a microscopic pictures for the heat
conduction\cite{hea} and thermal diffusion\cite{dif} by using the thermo
field
dynamics\cite{tft}. However, the calculations got so involved as to obscure
the picture. We are led to try a simple model, if not quite realistic,
to see in a mathematically transparent way
that an irreversibility can result from a reversible equation when the
initial condition involves temperature inhomogeneity. In this calculation,
we
use the standard machinery of statistical mechanics.

After a brief introduction to the thermo field dynamics in Sec.2,
we shall relate in Sec.3 the formulations of the state of local equilibrium
under temperature gradient in the thermo field dynamics and in the
more usual statistical mechanics. It is interesting that the thermo
field dynamics exhibits a non-local feature, though microscopic, of the
local equilibrium states; statistical mechanics gives an equivalent
description but in a more indirect way. Then in Sec. 4, we employ a simple
model to show in a mathematically transparent way that a reversible equation
of motion with the initial condition having inhomogeneous temperature can
lead to irreversibility. Our result is
irreversible in that it shows a diffusive behaviour among other
features, and yet symmetric with respect to the direction of time in
that the diffusion towards future/past comes out when the equation is
solved towards future/past, respectively.

\section {THERMO FIELD DYNAMICS}

The basic idea of the thermo field dynamics is to represent the thermal
state not by the mixture state with the density matrix $\rho(\beta)$ but by
a
pure state $|\beta \rangle$ which can be obtained from $\rho(\beta)$ by the
GNS construction. Thus, the thermal average of an observable $A$ reads
\begin{equation}
\Tr \, [\rho(\beta) A] = \langle \beta| A | \beta \rangle ,
\end{equation}
the left-hand side belonging to the usual statistical mechanics and
the right to the thermo field dynamics.

With any complete set $\{|n \rangle \}$ of orthonormal states, one can write
\begin{equation}
|\beta \rangle := \sum_n \rho(\beta)^{1/2} |n \rangle \otimes \langle n| .
\end{equation}
In fact
\begin{eqnarray*}
\langle \beta| A | \beta \rangle &=& \sum_m \sum_n
\langle m |\rho(\beta)^{1/2} A \rho(\beta)^{1/2} | n\rangle
\langle n| m \rangle \\
&=& \sum_n \langle n | \rho A | n \rangle = \Tr \, [\rho(\beta) A] .
\end{eqnarray*}

The thermo field dynamics further introduces operators that act on the bra
part as
\begin{equation}
(I \otimes A) |\psi \rangle \otimes \langle \chi | := |\psi \rangle \otimes
\Bigl(A^{\dagger} |\chi \rangle \Bigr)^{\rm t},
\end{equation}
thus doubling the number of operators to be considered. One uses shorthand
notations,
\begin{equation}
\widetilde{A}^{\dagger} := I \otimes A, \qquad |\psi \rangle \langle \chi |
:= |\psi \rangle \otimes \langle \chi | .
\end{equation}
Then, one finds: $\widetilde{c A} = c^{*} \widetilde{A}$, and
\begin{equation}
[\alpha, \alpha^{\dagger}] = 1 \quad \Longrightarrow \quad
[\widetilde{\alpha}, \widetilde{\alpha}^{\dagger}] = 1 , \quad {\rm etc.}.
\end{equation}

Suppose $\alpha$ acts on $|n \rangle$ as an annihilation operator,
$$
\alpha |n \rangle = \sqrt{n} |n - 1 \rangle
$$
and $\rho = Z_1^{-1} e^{-\beta {\cal H}}$ with ${\cal H} =
\omega \alpha^{\dagger} \alpha$, so that
\begin{equation}
|\beta \rangle = Z_1(\beta)^{-1/2} \sum_n e^{-n \beta \omega/2}|n \rangle
\langle n|,
\end{equation}
then
\begin{eqnarray*}
\alpha |\beta \rangle &=& \frac{1}{Z_1(\beta)^{1/2}}
\sum_{n=1}^{\infty} e^{- n \beta \omega/2}
\sqrt{n} |n-1 \rangle \langle n | \\
&=& \frac{e^{-\beta \omega/2}}{Z_1(\beta)^{1/2}}
\sum_{n=0}^{\infty} e^{-n \beta \omega/2}
\sqrt{n + 1} |n \rangle \langle n + 1| , \\
\widetilde{\alpha}^{\dagger} |\beta \rangle &=&
\frac{1}{Z_1(\beta)^{1/2}} \sum_{n=0}^{\infty} e^{- n \beta \omega/2}
\sqrt{n + 1} |n \rangle \langle n +1 | ,
\end{eqnarray*}
implying
$$
\Bigl(\alpha - e^{-\beta \omega/2} \widetilde{\alpha}^{\dagger}\Bigr)
|\beta \rangle = 0 ,
$$
which is called the thermal state condition.  In terms of $\theta$ as
defined for the $\beta$ by $\tanh \theta = e^{-\beta \omega/2}$,
\begin{equation}\label{eq:Bog}
\Bigl(\alpha \cosh \theta - \widetilde{\alpha}^{\dagger} \sinh \theta\Bigr)
|\beta \rangle = 0 ,
\end{equation}
which reminds us of the Bogoliubov transformation\cite{bog}:
\begin{equation}\label{eq:Bog2}
U(\beta) \alpha U(\beta)^{-1} = \alpha \cosh \theta -
\widetilde{\alpha}^{\dagger} \sinh \theta,
\end{equation}
with
$$
U(\beta) = \exp \Bigl[ \Bigl(\alpha^{\dagger}\widetilde{\alpha}^{\dagger} -
\alpha \widetilde{\alpha}\Bigr) \theta \Bigr] .
$$
This operator generates the state $|\beta \rangle$ from
the ``no particle state'' $|\Omega \rangle := |0 \rangle \langle 0|$ of
$\alpha$,
\begin{equation}\label{eq:U0}
U(\beta) |\Omega \rangle = |\beta \rangle
\end{equation}
as one sees
from the operator (\ref{eq:Bog2}) annhilating (\ref{eq:U0}), or from
\begin{eqnarray}
U(\beta) |\Omega \rangle &=& \frac{1}{Z_1(\beta)^{1/2}}
\exp [\alpha^{\dagger} \widetilde{\alpha}^{\dagger}
\tanh \theta] |\Omega \rangle \nonumber \\
&=& \frac{1}{Z_1(\beta)^{1/2}} \sum_{n=0}^{\infty}
e^{- n\beta \omega/2} |n \rangle \langle n| = |\beta \rangle .
\end{eqnarray}

We have so far considered only a single mode.
In general, however, the object has many modes, and often infinitely many,
to which a tensor product of the states considered above corresponds.

\section{NONUNIFORM TEMPERATURE}

Suppose we change the temperature a bit uniformly all over the system. Then,
\begin{eqnarray*}
|\beta + \mit\Delta \beta \rangle &=& U(\beta + \mit\Delta \beta)
|\Omega \rangle  \\
&=& \frac{1}{{Z_1}(\beta + \mit\Delta \beta)^{1/2}}
\exp \Bigl[\alpha^{\dagger} \widetilde{\alpha}^{\dagger}
e^{-(\beta + \mit\Delta \beta) \omega/2}\Bigr] |\Omega \rangle,
\end{eqnarray*}
which can be approximated as
\begin{eqnarray*}
|\beta + \mit\Delta \beta \rangle = N_1 \left( 1 -
\frac{\omega \mit\Delta \beta}{2}
e^{- \beta \omega/2} \alpha^{\dagger} \widetilde{\alpha}^{\dagger}
\right) |\beta \rangle
\end{eqnarray*}
with $N = Z_1(\beta)^{1/2}/Z_1(\beta + \mit\Delta \beta)^{1/2}$,
the normalization factor.

For a system with many modes labelled by ${\bvec k}$, $Z_1$ is replaced by
$Z$, and
\begin{equation}
|\beta + \mit\Delta \beta \rangle = N (1 - i\Gamma) |\beta \rangle
\end{equation}
with
$$
\Gamma = -i \sum_{\rm k} \frac{\omega_{\rm k} \mit\Delta \beta }{2}
e^{- \beta \omega_{\rm k}/2} \alpha_{\rm k}\,^{\dagger}
\widetilde{\alpha_{\rm k}}^{\dagger} .
$$
Define:
\begin{equation}
\phi({\bvec x}) := \frac{1}{\sqrt{V}}\sum_{\rm k} \alpha_{\rm k}
e^{i{\rm k}\cdot{\rm x}}, \qquad
\widetilde{\phi}({\bvec x}) := \frac{1}{\sqrt{V}}\sum_{\rm k}
\widetilde{\alpha}_{\rm k} e^{i{\rm k}\cdot{\rm x}},
\end{equation}
then
\begin{equation}\label{eq:G}
G = - i \mit\Delta \beta \int
\phi^{\dagger}\left({\bvec x} + \frac{{\bvec \xi}}{2}\right)
g({\bvec \xi})
\widetilde{\phi}^{\dagger}\left({\bvec x} - \frac{{\bvec \xi}}{2}\right)
d^3 x d^3 \xi .
\end{equation}

For a gas of free particles (or quasi-particles), for which
\begin{equation}
\omega_{\rm k} = \frac{k^2}{2m},
\end{equation}
one has, in high temperature approximation,
\begin{equation}
g({\bvec \xi}) = \left(\frac{m^3}{\pi^3 \beta^5 \hbar^{6}}
\right)^{1/2}
\left(3 - \frac{2m}{\beta \hbar^2}\xi^2\right)
\exp \left[- \frac{m}{\beta \hbar^2} \xi^2 \right] ,
\end{equation}
which is short-ranged as shown in Table 1.
\begin{center}
{\bf Table 1}@The range of $g$.

\begin{tabular}{r|cccc} \hline \hline
     &       &         &       &          \\
$m$  & \quad & nucleon & \quad & electron \\ \hline
     &       &         &       &          \\
$\displaystyle{\sqrt{\frac{\beta \hbar^2}{m}} = T ({\rm K})^{-1/2}
\times}$ & & $7 \times 10^{-10}$ m & & $3 \times 10^{-8}$ m \\
         & &                       & &         \\ \hline
\end{tabular}
\end{center}

Since $g({\bvec \xi})$ is short-ranged, the operator (\ref{eq:G}) can be
regarded as
a sum of local operators. Therefore, the state with {\it nonuniform}
temperature $\beta + \beta_a({\bvec x})$ may be represented as
\begin{equation}\label{eq:beta}
|\beta + \beta_a \rangle = N \Bigl(1 - i \Gamma[\beta_a] \Bigr)
|\beta \rangle
\end{equation}
with the temperature inhomogeneity $\beta_a({\bvec x})$ being pushed into
the integrand of (\ref{eq:G}):
\begin{equation}\label{eq:tft}
\Gamma [\beta_a] = -i \int
\phi^{\dagger}\left({\bvec x} + \frac{{\bvec \xi}}{2}\right)
\widetilde{\phi}^{\dagger}\left({\bvec x} - \frac{{\bvec \xi}}{2}\right)
g({\bvec \xi}) \beta_a({\bvec x}) d^3 x d^3 \xi ,
\end{equation}
since the temperature inhomogeneity $\beta_a({\bvec x})$ we
are concerned with in studying thermal processes varies only slowly on a
macroscopic scale.

We add that (\ref{eq:G}) is often written as
\begin{equation}
\Gamma[\beta_a] =  \int
\Phi^{\dagger}\left({\bvec x} + \frac{{\bvec \xi}}{2}\right) \tau_2
\widetilde{\Phi}\left({\bvec x} - \frac{{\bvec \xi}}{2}\right)
g({\bvec \xi}) \beta_a({\bvec x}) d^3 x d^3 \xi .
\end{equation}
in terms of the thermal doublet,
\begin{equation}
\Phi({\bvec x}) := \left(
\begin{array}{c}
\phi_1({\bvec x}) \\
\phi_2({\bvec x})
\end{array} \right) = \frac{1}{\sqrt{V}} \left(
\begin{array}{c}
a_{\rm k} \\
\widetilde{a}_{\rm k}\,^{\dagger}
\end{array}
\right) e^{i{\rm k}\cdot{\rm x}} .
\end{equation}
This is the representation we have been using for the state with nonuniform
temperature in our studies of heat conduction and thermal diffusion
\cite{hea}\cite{dif}.

As is clear from its derivation, this representation is equivalent to the
one by the density matrix,
\begin{equation}\label{eq:rhob}
\rho[\beta + \beta_a({\bvec x})] := \frac{1}{Z}
\exp \left[- \beta {\cal H} -
\int \beta_a({\bvec x}) {\cal H}({\bvec x}) d^3x \right] ,
\end{equation}
or, more precisely, to its expansion to the first order in $\beta_a$,
\begin{eqnarray}\label{eq:stm}
\lefteqn{\rho[\beta + \beta_a({\bvec x})]
= Z^{-1} e^{-\beta {\cal H}}}\nonumber \\
&&\times \left(1 -
\int_0^1 d \lambda e^{\lambda \beta{\cal H}}\int d^3 x \beta_a({\bvec x})
{\cal H}(\bvec x) e^{-\lambda \beta {\cal H}}\right) ,
\end{eqnarray}
provided that $\beta_a({\bvec x})$ is slowly varying.

It is interesting to compare (\ref{eq:stm}) and (\ref{eq:beta}) +
(\ref{eq:tft}) observing that the effect of the temperature
inhomogeneity $\beta_a({\bvec x})$ appears not to be strictly local in
thermo field dynamics but has a small extention as represented by
$g({\bvec \xi})$. We repeat that the two repesentations are equivalent,
and the effect of $g({\bvec x})$ is also there in (\ref{eq:stm})
as a little analysis shows. It is only
that the effect can be seen more easily in the thermo field dynamics.

Another remark concerns Zubarev's assertion \cite{Zub} that no 
irreversible processes can follow from the density matrix 
(\ref{eq:rhob}).  Our model calculation in the next section will 
provide a counterexample to his assertion. Further discussion will 
be given in Sec. V.

\section{A MODEL EXHIBITING IRREVERSIBILITY}

To see that irreversibility can result from
the interaction of particles even when their equation of motion is
reversible, we consider a simple
model\cite{lee}\cite{hua} of a Bose gas (particle mass: 1/2) with a
quadratic Hamiltonian,
\begin{equation}\label{eq:K}
{\cal H} :=
\sum_{\rm p} p^2 a_{\rm p}^{\dagger} a_{\rm p} +
\frac{\rho_0}{2} \sum_{p \le K} v_0 ( a_{\rm p}^{\dagger}
a_{-{\rm p}}^{\dagger}
+ 2  a_{\rm p}^{\dagger} a_{\rm p} +  a_{\rm p}  a_{-{\rm p}}) .
\end{equation}
This Hamiltonian is time-reversible, as one sees by applying time
reversal ${\bvec p} \to -{\bvec p}$ to the momenta that label the
operators $a_{\rm p}$.  The price we have
to pay for the simplicity of its analysis is the non-conservation
of the particle number; this is serious because we are going to
look at the particle diffusion as a sign of irreversibility.
Nevertheless, we shall be able to see this sign \cite{vic}, though
superposed with an unwanted decay of the particle number.

We use the representation (\ref{eq:rhob}) of the
temperature inhomogeneity in terms of the density matrix, for
which we need the Hamiltonian density,

\begin{eqnarray*}
{\cal H}({\bvec x}) &=&
\frac{\partial  \phi^{\dagger}({\bvec x})}{\partial x_l}
\frac{\partial  \phi({\bvec x})}{\partial x_l} \\
&&\!\!\!\!+ \frac{\rho_0}{2}
\int v({\bvec r})\Bigl[
\phi^{\dagger}({\bvec x}+\frac{1}{2}{\bvec r})
\phi^{\dagger}({\bvec x}-\frac{1}{2}{\bvec r})  \\
& &\!\!\!\! +  \phi^{\dagger}({\bvec x}+\frac{1}{2}{\bvec r})
\phi({\bvec x}-\frac{1}{2}{\bvec r}) +
\phi^{\dagger}({\bvec x}-\frac{1}{2}{\bvec r})
\phi({\bvec x}+\frac{1}{2}{\bvec r}) \\
&&\!\!\!\! +  \phi({\bvec x}+\frac{1}{2}{\bvec r})
\phi({\bvec x}-\frac{1}{2}{\bvec r}) \Bigr] d^3 {\rm r}
\end{eqnarray*}
where $v({\bvec r})$ is such that
\begin{equation}
\tilde v({\bvec k}) =
\int v({\bvec r}) e^{-i{\rm k}\cdot{\rm r}} d^3{\rm r} = \left\{
\begin{array}{ccl}
v_0 &  &(k \le K) \\
0   &  & (k > K),
\end{array} \right.
\end{equation}
or
$$
v({\bvec r}) = 4\pi^2 v_0 K^2 j_1(Kr)/r .
$$
In terms of the Fourier transform, we find
\begin{eqnarray*}
\lefteqn{\int v({\bvec r})  \phi^{\dagger}({\bvec x} + \frac{1}{2}{\bvec r})
\phi({\bvec x} - \frac{1}{2}{\bvec r}) d^3{\rm r}}\nonumber \\
&& \qquad =
\frac{1}{V} \sum_{{\rm p}'{\rm p}} \tilde v(\frac{{\bvec p}' + {\bvec
p}}{2})
a^{\dagger}_{{\rm p}'} a_{\rm p}
e^{i({\rm p}-{\rm p}')\cdot {\rm x}}
\end{eqnarray*}
and similarly for their Hermitian relatives. We assume for simplicity
that $K$ is sufficiently
large to make all the $\tilde v({\bvec k})$'s for
thermally agitated particles to have the constant value $v_0$; we
shall hereafter suppress the proviso $p \le K$ .

For the particle number density,
\begin{equation}
j_0({\bvec x}) = \phi^{\dagger}({\bvec x})\phi({\bvec x}) =
\frac{1}{V}\sum_{\rm p, p'}a_{\rm p'}^{\dagger} a_{\rm p}
e^{i({\rm p} - {\rm p}')\cdot {\rm x}},
\end{equation}
we wish to find out the long-time behaviour of the deviation,
\begin{eqnarray}\label{eq:DO1}
\lefteqn{\langle \hat j_0({\bvec x}, t) \rangle^{(1)} }\nonumber \\
&&= - \frac{1}{Z^{(0)}}\Tr \left[e^{-\beta {\cal H}}
\int_0^1 d\lambda \int d^3 x'
\beta_a({\bvec x}') {\cal H}({\bvec x}')\right.\nonumber\\
&&\quad \left. \times e^{i {\cal H}\tau}
j_0({\bvec x}) e^{-i {\cal H}\tau} \right] ,
\end{eqnarray}
from the average,
\begin{equation}
\langle \hat j_0({\bvec x}, t) \rangle^{(0)} =
\frac{1}{Z^{(0)}} \Tr \, [e^{-\beta {\cal H}}j_0({\bvec x})]
\end{equation}
where
$$
\tau = t + i\lambda \beta
$$
and we have assumed that
$$
\int \beta_a({\bvec x}) d^3 x = 0.
$$

The Hamiltonian (\ref{eq:K}) can be diagonalized as
\begin{equation}
{\cal H} = \sum_{\rm p} \omega_p
\alpha_{\rm p}^{\dagger} \alpha_{\rm p} \qquad
\bigl(\omega_p := p \sqrt{p^2 + 2 c^2}, \quad c^2 := \rho_0 v_0 \bigr) ,
\end{equation}
where the zero-point energy has been dropped, by a Bogoliubov transformation
$$
a_{\rm p} = \alpha_{\rm p} \cosh \theta_p -
\alpha_{-\rm p}^{\dagger} \sinh \theta_p ,
$$
with
\begin{equation}
\begin{array}{rcl}
\sinh^2 \theta_p &=& \displaystyle{\frac{p^2 +
c^2}{2 p\sqrt{p^2 + 2c^2}} - \frac{1}{2}}, \\
\cosh \theta_p \sinh \theta_p &=& \displaystyle{
\frac{c^2}{2p \sqrt{p^2 + 2 c^2}}} .
\end{array}
\end{equation}

The thermal average (\ref{eq:DO1}) turns out to be given by the sum:
\begin{eqnarray*}
\langle j_0({\bvec x}, t)\rangle^{(1)} &=&
\langle j_0({\bvec x}, t)\rangle^{({\rm s},-)} +
\langle j_0({\bvec x}, t)\rangle^{({\rm s},+)}\\
& & + \langle j_0({\bvec x}, t)\rangle^{({\rm c},-)} +
\langle j_0({\bvec x}, t)\rangle^{({\rm c},+)},
\end{eqnarray*}
where
\begin{eqnarray}\label{eq:j+pm}
\lefteqn{\langle j_0({\bvec x}, t) \rangle^{(\kappa,\pm)}}\nonumber\\
&& = \frac{-1}{2(2\pi)^6}
\int d^3{\rm k} \,
\beta_a({\bvec k}) e^{i{\rm k}\cdot{\rm x}} \int_0^1
F^{(\kappa, \pm)}({\bvec k}, \lambda)
\end{eqnarray}
with
\begin{eqnarray}
F^{({\rm s}, \pm)}({\bvec k}, \tau) &=&
\int d^3{\rm P} \, g({\bvec p}',{\bvec p}) \sinh(\theta_{p'} + \theta_p)
\nonumber \\
& &\!\!\!\!\!\!\!\!\!\!\!\!\!\! \times  \left\{
\begin{array}{l}
f(\omega_{p'})f(\omega_p) e^{-i(\omega_{p'} + \omega_p)\tau} \\
\bigl\{f(\omega_{p'}) + 1\bigr\} \bigl(f(\omega_p)+1\bigr) e^{i(\omega_{p'}
+
\omega_p)\tau}
\end{array}\right.
\end{eqnarray}
and
\begin{eqnarray}\label{eq:Fc}
F^{({\rm c},\pm)}({\bvec k}, \tau) &=&
\int d^3{\rm P} \, h({\bvec p}',{\bvec p}) \cosh(\theta_{p'} + \theta_p)
\nonumber \\
& & \times \left\{
\begin{array}{l}
f(\omega_{p'})\bigl\{f(\omega_p)+1\bigr\} e^{-i(\omega_{p'} - \omega_p)\tau}
\\
\bigl\{f(\omega_{p'}) + 1 \bigr\} f(\omega_p)
e^{i(\omega_{p'} - \omega_p)\tau} ,
\end{array}\right.
\end{eqnarray}
Here,
${\bvec p}' = {\bvec P} + {\bvec k}/2$,
${\bvec p} = {\bvec P} - {\bvec k}/2$,
and
\begin{eqnarray}\label{eq:gh}
\lefteqn{\left.
\begin{array}{c}
g({\bvec p}', {\bvec p}) \\
h({\bvec p}', {\bvec p})
\end{array} \right\}
:= ({\bvec p}'\cdot{\bvec p}) \left\{
\begin{array}{c}
\sinh \\
\cosh
\end{array}\right\}(\theta_{p'} + \theta_p)}  \nonumber \\
& & \mp
c^2 (\cosh \theta_{p'} - \sinh \theta_{p'})
(\cosh \theta_{p} - \sinh \theta_p),
\end{eqnarray}
of which $g$ is associated with the ``off-diagonal'' term
$\alpha_{{\rm p}'}^{\dagger}\alpha_{-{\rm p}}^{\dagger}$
and $h$ with the ``diagonal''
$\alpha_{{\rm p}'}^{\dagger}\alpha_{{\rm p}}$ in the Hamiltonian
density $\hat {\cal H}({\bvec x})$ . We note
\begin{eqnarray}\label{eq:Tra}
\lefteqn{\frac{1}{Z^{(0)}}\Tr\Bigl[e^{-\beta {\cal H}}
\alpha_{{\rm p}'}^{\dagger}\alpha_{-{\rm p}}^{\dagger}
\alpha_{-{\rm q}'}\alpha_{{\rm q}}\Bigr]}\nonumber \\
& & = f(\omega_{p'})f(\omega_p)
(\delta_{{\rm p}',-{\rm q}'}\delta_{-{\rm p},{\rm q}} +
\delta_{{\rm p}',{\rm q}} \delta_{{\rm p},{\rm q}'}) .
\end{eqnarray}

\subsection{Evaluation of the integral $F^{({\rm s},\pm)}$}

As a key for the following calculations, the temperature inhomogeneity
$\beta_a({\bvec x})$ is assumed to be varying on a macroscopic
scale, so that the support of its Fourier transform is very small
in comparison with the scales on which $\omega_p$ and the coefficients of
the Bogoliubov transformation vary.

We begin the evaluation of (\ref{eq:j+pm}) from the ${\bvec P}$-integration,
\begin{eqnarray}\label{eq:Omeg}
F^{({\rm s},+)}({\bvec k}, \tau) &=&
\int_0^{\infty} P^2 d P \, \sinh [2 \theta_{P}] \Bigl(f(\omega_P)\Bigr)^2  
\nonumber
\\
& & \times \int d\Omega_{\rm P}
g({\bvec p}',{\bvec p})  e^{-i(\omega_{p'} + \omega_p)\tau},
\end{eqnarray}
where, in sinh and $f$,
we have put ${\bvec p} = {\bvec p}'= {\bvec P}$ taking advantage of
the small support of $\tilde \beta({\bvec k})$. However, we cannot do so
in $g$ and the
exponent, because $g$ would vanish if we did and the exponent is
proportional
to $\tau$ that involves the macroscopically large variable $t$ .
Instead, we expand to order $k^2$,
\begin{equation}\label{eq:vs}
\omega_{p'} + \omega_{p} = 2 \omega_p + \frac{P^2 + c^2}{2\omega_p} k^2
- \frac{c^4}{\omega_p^3} ({\bvec P}\cdot {\bvec k})^2
\end{equation}
and put
\begin{eqnarray*}
g({\bvec p}',{\bvec p}) &=& g_0(P) + g_1(P)({\bvec P}\cdot{\bvec k})\\
& & + g_2(P){\bvec k}\cdot{\bvec k} + g_3(P)
\frac{({\bvec P}\cdot{\bvec k})^2}{c^2} ;
\end{eqnarray*}
finding, however, that $g_0 = g({\bvec P}, {\bvec P}) =
0$ by the Bogoliubov transformation
diagonalizing ${\cal H}$ [{\it cf.} the remark after (\ref{eq:gh})],
$g_1 = 0$ by the symmetry $g({\bvec p}',{\bvec p}) = g({\bvec p},{\bvec
p}')$
from (\ref{eq:gh});  we take in advance the later result  into account that
$$
g_3 (P) = \cosh 2\theta_P \sinh^2 2\theta_P
$$
vanishes at the value of $P$ that concerns us.
Thus,
\begin{equation}\label{eq:gex}
g({\bvec p}', {\bvec p}) = g_2(P) \quad \mbox{with} \quad
g_2(P) = -\frac{1}{2} \sinh 2\theta_P .
\end{equation}

The integration over the angles in (\ref{eq:Omeg}),
$$
I_0 = \int_0^{\pi} e^{i\alpha \cos^2 \gamma} \sin \gamma d\gamma
= \frac{2}{\sigma} \int_0^{\sigma} e^{-y^2} dy,
$$
is carried out by adding and subtracting 
$\displaystyle{\frac{2}{\sigma}\int_{\sigma}^{\infty}
e^{- y^2} dy}$ taking advantage of
Im $\alpha = \lambda \beta > 0$,
where $\gamma$ is the angle between ${\bvec P}$ and ${\bvec k}$,
\begin{equation}
\alpha = \frac{c^4 P^2}{\omega_P^3}k^2 \tau, \quad
\sigma := (- i\alpha)^{1/2}, \quad \mbox{and} \quad y = \sigma x .
\end{equation}
We note $- \pi/2 < \arg \sigma < 0$ and $|\sigma| \to \infty$ as
$t \to \infty$. Then, asymptotically
\begin{equation}
I_0 \sim \frac{\sqrt{\pi}}{\sigma} - \frac{1}{\sigma^2}e^{-\sigma^2} .
\end{equation}
Thus,
\begin{eqnarray}\label{eq:F+}
\lefteqn{
F^{({\rm s},+)}({\bvec k},\tau)}\nonumber\\
&& \sim
2\pi \int_0^{\infty}P^2 dP\, \sinh [2\theta_P]
\Bigl(f(\omega_P)\Bigr)^2  g_2(P) k^2 \nonumber \\
&& \times \left\{\frac{\sqrt{\pi}}{\sigma} -\frac{1}{\sigma^2}
\exp\left[i\frac{c^4 P^2}{\omega_P^3}k^2 \tau \right] \right\}\nonumber \\
&& \times \exp\left[-i\left(2\omega_P +
\frac{P^2 + c^2}{2\omega_P}k^2\right)\tau \right] .
\end{eqnarray}

Similar calculations give
\begin{eqnarray}\label{eq:F-}
\lefteqn{
F^{({\rm s},-)}({\bvec k},\tau)}\nonumber \\
&& \sim 2 \pi
\int_0^{\infty}P^2 dP\, \sinh [2\theta_P]
\Bigl(f(\omega_P) + 1 \Bigr)^2 g_2(P)k^2 \nonumber \\
&& \times \left\{i \frac{\sqrt{\pi}}{\sigma} + \frac{1}{\sigma^2}
\exp\left[- i\frac{c^4 P^2}{\omega_P^3}k^2 \tau \right] \right\}
\nonumber \\
& & \times \exp\left[i\left(2\omega_P +
\frac{P^2 + c^2}{2\omega_P}k^2\right)\tau
\right] .
\end{eqnarray}
If we change the variable of integration $P$ to $-P$ in (\ref{eq:F-})
after analytically continuing the integrand
to the complex $P$-plane with two cuts to cope with the singularity of
$(P^2 + 2c^2)^{1/2}$; one extending along the imaginary axis
from $i\sqrt{2} c$ to ${i\infty}$, and the other from
$-i\sqrt{2} c$ to ${-i \infty}$, then $\omega_{- P} = - \omega_{P}$, and
consequently
\begin{eqnarray}
f(\omega_{- P}) + 1 &=&
\frac{e^{- \beta \hbar \omega_P }}{e^{- \beta\hbar \omega_P} - 1} =
- f(\omega_{P}), \nonumber \\
\sinh \theta_{-P} &=& - \sinh \theta_P
\end{eqnarray}
so that for $F^{({\rm s})} := F^{({\rm s},+)} + F^{({\rm s},-)}$,
\begin{eqnarray}\label{eq:1A2A}
\lefteqn{
F^{({\rm s})}({\bvec k},\tau) }\nonumber \\
&& \sim \int_{-\infty}^{\infty}dP\, G(P)
\left\{\frac{\sqrt{\pi}}{\sigma} -\frac{1}{\sigma^2}
\exp\left[i\frac{c^4 P^2}{\omega_P^3}k^2 \tau \right]\right\}\nonumber \\
& & \times \exp\left[-i\left(2\omega_P + \frac{P^2 +
c^2}{2\omega_P}k^2\right)
\tau \right] ,
\end{eqnarray}
where
$$
G(P) := 2 \pi P^2 \sinh [2\theta_P] \bigl\{f(\omega_P)\bigr\}^2 g_2(P)\,
k^2.
$$

\subsubsection{Long-time behavior of $F^{(\rm s)}_1$}

The time scale we are interested in is macroscopic.
To evaluate the integral (\ref{eq:1A2A}) asymptotically for $t \to \infty$,
we use the saddle-point method.
Candidates for the saddle point $P_{{\rm s}, 1}$ for the part of the
integral
(\ref{eq:1A2A}),
\begin{eqnarray}
\lefteqn{
J_1 := - \frac{1}{\sigma^2} \int_{-\infty}^{\infty} dP \, G(P) }\nonumber \\
& & \times \exp\left[-i\left\{2\omega_P + \left(\frac{P^2 + c^2}{2\omega_P}
- \frac{c^4 P^2}{\omega_P^3}\right)k^2 \right\} \tau \right],
\end{eqnarray}
are determined by the condition of vanishing derivative of the exponent.
\begin{eqnarray*}
2 \frac{d\omega_P}{dP} - \left\{
\left(\frac{P^2 + c^2}{2\omega_{P}^2} \right. \right. &-& \left.
\frac{3c^4P^2}{\omega_P^4}\right)
\frac{d\omega_{P}}{dP}  \\
& + & \left. \left(\frac{P}{\omega_P} -
\frac{2c^4 P}{\omega_{P}^3}\right) \right\} k^2 = 0,
\end{eqnarray*}
where
$$
\frac{d \omega_P}{dP} = \frac{2(P^2 + c^2)}{\sqrt{P^2 + 2c^2}}.
$$
For small ${\bvec k}$, the main part of the root is given by $P = \pm ic$.
Then, we put $P^2 = - c^2 + \xi$, finding $\xi = - (3/4) k^2$
up to order $k^2$, and hence  $P_{{\rm s},1} =
\pm i c \bigl\{1 + (3/8)(k^2/c^2)\bigr\}$ as  candidates
for the saddle points. Since
$$
\frac{d^2\omega_P}{dP^2} = \frac{2 P(P^2 + 3c^2)}{(P^2 + 2c^2)^{3/2}}
$$
takes the value $\pm 4i + O(k^2/c^2)$ at the saddle points, the
exponent behaves like
\begin{eqnarray*}
\lefteqn{2\omega_P + \left(\frac{P^2 + c^2}{2\omega_P}
- \frac{c^4 P^2}{\omega_P^3}\right) k^2} \\
& & = \pm i \Big\{(2c^2 + k^2) + 4 (P - P_{{\rm s},1})^2\Big\}
\end{eqnarray*}
in the neighborhod of $P_{{\rm s},1}$, so that
\begin{equation}
J_1 \sim - \frac{1}{\sigma^2}e^{\pm(2 c^2 + k^2)\tau}
\int_{-\infty}^{\infty} G(P) e^{\pm 4(P - P_{{\rm s},1})^2 \tau}
dP .
\end{equation}
For $\tau \sim t \to +\infty$, therefore, we should take the lower sign 
here
\begin{equation}\label{eq:sadd}
P_{{\rm s}, 1} = - ic \left(1 + \frac{1}{8}\frac{k^2}{c^2}\right)
\end{equation}
for the saddle point which the path of integration should pass.  Thus,
\begin{equation}\label{eq:sdot}
F^{({\rm s})} \sim - \frac{1}{\sigma(P)^2} G(P_{{\rm s},1})
e^{-(2 c^2 + k^2) \tau }\int_{-\infty}^{\infty}
e^{- 4 \tau s^2} ds,
\end{equation}
and we have only to carry out the expansion (\ref{eq:gex}) around
$P = P_{{\rm s},1}$ , obtaining
%
\begin{equation}\label{eq:F1}
F^{({\rm s})}({\bvec k}, \tau)_1 \sim - \frac{1}{2}
\left(\frac{\pi}{\tau}\right)^{3/2}f(\omega_{P_{{\rm s},1}})^2
e^{-(2c^2 + k^2)\tau}.
\end{equation}
Note that the small factor $k^2$ of $G$ has been cancelled by $1/\sigma^2$.
The factor $e^{- k^2 \tau}$ is a welcome sign of the diffusive behavior.
However, it is accompanied by $e^{- 2 c^2\tau}$, a damping, which
is an unexpected feature.

\subsubsection{Long-time behavior of $F^{(\rm s)}_2$}

We still have to evaluate the other part of the integral (\ref{eq:1A2A}),
\begin{eqnarray}\label{eq:J2}
\lefteqn{J_2 := \frac{\sqrt{\pi}}{\sigma} \int_{-\infty}^{\infty}
dP \, G(P)}\nonumber \\
&& \times
\exp\left[-i\left(2\omega_P + \frac{P^2 + c^2}{2\omega_P}k^2\right)\tau
\right] .
\end{eqnarray}
Candidates for the saddle point
$P_{{\rm s}, 2}$
for this integral are given by the roots of
$$
2 \frac{d\omega_P}{dP} -
\left(\frac{P^2 + c^2}{2\omega_{P}^2} \frac{d\omega_{P}}{dP} -
\frac{P}{\omega_P}\right) k^2 = 0,
$$
and hence
\begin{equation}\label{eq:sad}
P_{{\rm s},2} = \pm ic\left(1 + \frac{1}{8}\frac{k^2}{c^2} \right)
\end{equation}
up to order $k^2$. Since
$$
\omega_{P_{{\rm s},2}} = \pm i c \left(1 + \frac{k^2}{8c^2}\right)
\left(c ^2 - \frac{k^2}{4}\right)^{1/2} = \pm i c + O(k^4),
$$
we see the exponent in (\ref{eq:J2}) behaves as
$$
- i\left(2\omega_P + \frac{P^2 + c^2}{2\omega_P}k^2\right) \tau  =
\pm 2\big\{c^2 + 2 (P - P_{{\rm s},2})^2\bigr\} \tau
$$
in a neighbood of $P_{{\rm s},2}$. Therefore, we choose the saddle point
$\displaystyle{
P_{{\rm s},2} = - ic \left(1 + \frac{1}{8}\frac{k^2}{c^2} \right)
}$
to let the path of integration (\ref{eq:J2}) pass, obtaining
\begin{equation}\label{eq:F2}
F^{({\rm s})}({\bvec k}, \tau)_2 \sim \frac{1}{2} k
\left(\frac{\pi^2}{\tau}\right) f(\omega_{P_{{\rm s},1}})^2
e^{-2c^2 \tau},
\end{equation}
which implies a simple damping without any diffusive character.

\subsubsection{Integration over $\lambda$}

It remains to integrate (\ref{eq:F1}) and (\ref{eq:F2}) over $\lambda$.
We may consider only the part,
\begin{equation}
f(\omega_{P_{\rm s}})^2 \int_0^1 e^{-2 i c^2\beta\lambda} d\lambda =
- \frac{1}{2\beta c^2}\cot \frac{\beta c^2}{2}.
\end{equation}
Thus, summarizing
\begin{eqnarray}\label{eq:res1}
\lefteqn{
\langle j_0({\bvec x}, t)\rangle^{({\rm s})} = - \frac{1}{2(2\pi)^6}
\frac{1}{4\beta c^2} \cot \frac{\beta c^2}{2}}\nonumber \\
& & \times \left(\frac{\pi}{t}\right)^{3/2}  e^{- 2c^2 t}
\int \Bigl(e^{- k^2 t} - \sqrt{\pi k^2 t}\Bigr)
\beta_a({\bvec k}) d^3{\rm k} .
\end{eqnarray}

Two remarks are in order.  (1)  $e^{-k^2t}$ leads to a diffusion.
This does not contradict the basic reversibility of our model.
If we reversed the time to look at
the evolution towards the past, $t < 0,$  then we must take the path
of integration crossing the other saddle point as given by (\ref{eq:sad})
with the upper sign, obtaining the diffusion towards the past.  Our result
is time symmetric in this sense, yet exhibiting the irreversible
features. The same circumstance (but no damping) was found in our previous
approach\cite{dif} with the thermo field dynamics using a more realistic
model.

(2) The simple damping implied by (\ref{eq:F2})
should call for an interpretation if it could be taken as an irreversible
behavior. Its amplitude has a factor $k\sqrt{t}$ extra to (\ref{eq:F1}),
$k$ being small but $t$ large.  The damping might be an artifact due to
our model Hamiltonian not conserving the particle number.

\subsection{Evaluation of the integral (31)}

Since
\begin{eqnarray}
h({\bvec P}, {\bvec P}) = \omega_P, & &
\cosh[2\theta_P] = \frac{P^2 + c^2}{\omega_P}, \\
\omega_{{\rm P} + {\rm k}/2} - \omega_{{\rm P} + {\rm k}/2} &=&
2 \frac{P^2 + c^2}{\omega_P}({\bvec P}\cdot{\bvec k})
\end{eqnarray}
(\ref{eq:Fc}) turns out to be
\begin{eqnarray*}
\lefteqn{
F^{({\rm c},\pm)}= \int_0^{\infty}P^2 dP \,(P^2 + c^2)f(\omega_P)
\bigl\{f(\omega_P) + 1 \bigr\} }\nonumber \\
& & \times \int d\Omega_{\rm P}
\exp\left[\mp 2i \frac{P^2 + c^2}{\sqrt{P^2 + 2c^2}}k \tau \cos
\gamma\right] .
\end{eqnarray*}
The integration over angles gives the same result $F^{({\rm c})}/2$
for both $F^{({\rm c},\pm)}$:
\begin{eqnarray}\label{eq:Fcpm}
\lefteqn{
\frac{1}{2}F^{({\rm c})}}\nonumber \\
&&= \pi \int_0^{\infty}P^2 dP \,
\frac{\sqrt{P^2 + 2c^2}}{i k\tau}f(\omega_P)\bigl\{f(\omega_P) + 1 \bigr\}
\nonumber \\
& & \times \sum_{\pm} \pm \exp\left[\pm 2i \frac{P^2 + c^2}{\sqrt{P^2 +
2c^2}}
k \tau \right].
\end{eqnarray}

We use the saddle-point method after analytically continuing the integrand
on
the complex $P$-plane with two cuts, $[i\sqrt{2}c, \infty)$ and $(-\infty,
- i \sqrt{2}c]$ as before. The saddle points are
determined from the exponent by
\begin{equation}\label{eq:sadl}
\frac{d}{dP}\frac{P^2 + c^2}{(P^2 + 2c^2)^{1/2}} =
\frac{P(P^2 + 3c^2)}{(P^2 + 2c^2)^{3/2}} = 0
\end{equation}
as
$P_{\rm s} = \pm i\sqrt{3}c + \epsilon$, two twins, each of which,
being separated by one of the cuts, consists of the right and the left
member with an infinitesimal $\epsilon = + 0$ and  $\epsilon = - 0$,
respectively.
The other root $P = 0$ of the saddle point equation (\ref{eq:sadl})
is discarded because the prefactor vanishes there.
In the neighborhood of the saddle points $P_{\rm s}$, the exponent behaves
like
\begin{eqnarray*}
\lefteqn{
\frac{P^2 + c^2}{(P^2 + 2c^2)^{1/2}} =
\left.\frac{P^2 + c^2}{(P^2 + 2c^2)^{1/2}}
\right|_{P = \pm i\sqrt{3}c+\epsilon}}  \\
& & + \left.\frac{3c^4}{(P^2 + 2c^2)^{5/2}}\right|_{P = \pm
i\sqrt{3}c+\epsilon}(P - P_s)^2
\end{eqnarray*}
times $\pm 2i k\tau$. We note
$$
\Bigl.(P^2 + 2c^2)^{1/2}\Bigr|_{P = \pm i\sqrt{3}c + \epsilon} = \left\{
\begin{array}{ccl}
\pm i c & \quad & (\epsilon = + 0) \\
\mp i c &       & (\epsilon = - 0) .
\end{array}\right.
$$

Let us try the right members of the two twins, $P_{{\rm s}+} = \pm i
\sqrt{3}c
+ 0$. Then,
\begin{equation}\label{eq:sac}
\pm 2 i k \tau \frac{P^2 + c^2}{(P^2 + 2c^2)^{1/2}} = \mp 4ck \tau
\pm \frac{6 k \tau}{c}(P - P_{{\rm s} +})^2,
\end{equation}
which requires that the path of integration be deformed to run along the
imaginary axis and the quadrant of a large circle to come back to the
real axis. But, on the circle, the exponential behaves like
\begin{equation}\label{eq:dic}
\left|
\exp \left[\pm 2i k \tau \frac{P^2 + c^2}{(P^2 + 2c^2)^{1/2}}\right]
\right|
\sim \exp \bigl[ \mp 2 k \tau R \sin \phi \bigr]
\end{equation}
$(P := R e^{i\phi})$
requiring that $0 \le \phi \le \pi/2$ for $t > 0$. We remark that
$\bigl|f(\omega_P)\bigl\{f(\omega_P) + 1\bigr\}\bigr|$ behaves on the circle
as $e^{- R^2 |\cos 2 \phi|}$, not decreasing as $R \to \infty$ when
$\phi = \pm \pi/4$, while helping suppress (\ref{eq:dic}) when $\phi = 0,
\pi$.

Thus, we take the path of integration passing the saddle (\ref{eq:sac}),
$\bigl\{iy + 0: y \; \mbox{runs from} \; 0 \; \mbox{to} \; \pm R)
\bigr\} \cup \bigl\{R e^{i\phi}: \phi \; \mbox{runs from} \; \pm \pi/2 \;
\mbox{to} \; 0) \bigr\}$,
with $R \to \infty$, obtaining the same results for the two terms $\pm$ in
(\ref{eq:Fcpm}):
\begin{equation}\label{eq:Fc1}
F^{({\rm c})}= - \sqrt{\frac{3}{2}} c^{7/2}
\left(\frac{\pi}{k \tau}\right)^{3/2}f(\omega_{P_{\rm s}})
\bigl\{f(\omega_{P_{\rm s}}) + 1 \bigr\}e^{-4 c k \tau}.
\end{equation}
We remark that the double poles $P = \pm i \sqrt{2}c$ due to
$f(\omega_P)\bigl\{f(\omega_P) + 1\bigr\}$ have the residues
killed by the exponential.

In (\ref{eq:Fc1}), $\lambda$ appears only in $e^{- 4ck(t + i\lambda \beta)}$
and may be neglected because of the small factor $k$. Then,
\begin{eqnarray}
\lefteqn{
\langle j_0({\bvec x}, t)\rangle^{({\rm c})} = \frac{1}{2(2\pi)^6}
\sqrt{\frac{3}{2^5}}
\frac{c^{7/2}}{\displaystyle{\sinh^2 \frac{\sqrt{3}\beta c^2}{2}}}}
\nonumber \\
& & \times \int \beta_a({\bvec k})
\left(\frac{\pi}{k \tau}\right)^{3/2} e^{i{\rm k}\cdot{\rm x}-4ckt} d^3{\rm
k},
\end{eqnarray}
which is to be added on (\ref{eq:res1}).

Note that this result implies a kind of slow oscillatory diffusion as
illustrated by
\begin{eqnarray}\label{eq:osc}
\left(\frac{\pi}{k\tau}\right)^{3/2}
\int e^{i{\rm k}\cdot{\rm x} -4ck t } d^3{\rm k} &=&
\frac{4\pi^{5/2}}{|{\bvec x}|}
{\rm Im} \int_0^{\infty} \frac{e^{-(4ct - ix)k}}{\sqrt{k}} dk \nonumber \\
&=& \frac{8\pi^{5/2}}{|{\bvec x}|}
\frac{\sin \phi(|{\bvec x}|, t)}{\bigl\{x^2 + (4ct)^2\bigr\}^{1/4}},
\end{eqnarray}
where the sin of
$$
\phi(|{\bvec x}|, t) := \frac{1}{2} \tan^{-1}\frac{|{\bvec x}|}{4ct}
$$
is responsible for the oscillation mentioned.

Thus, we have seen that an irreversible behaviour can arise from the 
Hamiltonian dynamics with the initial condition having the 
temperature inhomogeneity represnted by the density matrix 
(\ref{eq:rhob}) and with the model Hamiltonian (\ref{eq:K}) which 
is time-reversal invariant.

\section{Discussion}

In this paper, we have first pointed out that the thermo field dynamics 
reveals a nonlocal structure hidden behind the local-equilibrium density 
matrix with temperature inhomogeneity in the standard statistical 
mechanics.

We have then shown with a simple model that, in a system of interacting 
particles, diffusive behaviour can arise from the Hamiltonian dynamics 
with an initial condition having temperature varying slowly in space 
even when the Hamiltonian is time-reversal invariant. The diffusive 
behaviour is seen when we look at the system in macroscopic scales both 
in time and space.  The macroscopic space scale is incorporated in our 
calculation by expansion in powers of the wave numbers, and the 
macroscopic time scale by applying the saddle point method to the 
integration over angular frequency.

It has to be remarked, however, that the diffusive behaviour 
(\ref{eq:res1}) we have obtained has in $x$-space the form, 
\begin{equation}\label{eq:diff}
\int t^{-3/2} e^{-m({\rm x} - {\rm y})^2/(\hbar t)}f({\bvec y}) d^3{\rm y},
\end{equation}
multiplied by a damping factor $e^{-2mc^2t/\hbar}$, if written fully with 
the particle mass $m$, the strength of the interaction $c$ and $\hbar$ 
restored. 

The damping just mentioned and the slow oscillatory diffusion we saw in 
(\ref{eq:osc}) are interesting new features. The latter may be a sound 
wave associated with the diffusion, which could have been smeared out to 
become invisible in macroscopic treatments, while the former could probably 
be an artifact due to our model Hamiltonian not conserving the particle number.

We notice further that the diffusion constant in (\ref{eq:diff}) is 
independent of the strength $c$ of the interaction against our expectation. 
This is a result of using expansions of different quantities in powers 
of the wave number $k$ and that, as is seen in (\ref{eq:sadd}) for 
example, many of the expansion coefficients have the interaction 
strength $c$ in their denominators, limiting the validity of our 
results, say (\ref{eq:diff}), to $k < c$ or $|{\bvec x} - {\bvec x}'| 
> 1/c$.  Therefore, we cannot let $c \to 0$ without losing the space 
region of validity of our results. If $c$ was 0, in fact, we would 
lose the diffusive behaviour because $\theta_p$ should vanish and 
consequently $F^{(s, \pm)} = 0$ in (\ref{eq:F+}) and (\ref{eq:F-}). 

After seeing all these features, we stress once again that we have shown 
by a simple example that the Hamiltonian dynamics with an initial 
condition having temperature inhomogeneity can lead to an irreversible 
behaviour even when the Hamiltonian is time-reversal invariant.  This is 
a fact we have established mathematically, though our model and some 
of its consequences are not quite physical.

The irreversibility so derived stands against Zubarev's assertion 
\cite{Zub} quoted at the end of Sec. III, but not entirely against 
his arguments because he argues simply that the results from 
Hamiltonian dynamics with a time-reversible Hamiltonian should be 
time-reversal symmetric. Our result is in fact symmetric under 
time reversal $t \to -t$ as remarked earlier. It is of course not 
symetric under time reversal with respect to any instant of time 
other than $t = 0$.  The instant $t = 0$ is special because the 
initial condition, which itself is time-symmetric, is given at this 
instant of time.\\

This work was presented at the International Workshop
on Thermal Field Theories and Their Applications
in Regensburg, Germany, Aug. 10 - 14, 1998.

\end{document}